%% file: gamma_chicj_final.tex
\RequirePackage{lineno}
\documentclass[prd,twocolumn,showpacs,amsmath,amssymb,nofootinbib]{revtex4-1}
\usepackage{epsfig}
\usepackage{dcolumn}
\usepackage{graphicx}
\usepackage{subfigure}
\usepackage{times}
\usepackage{setspace}
\usepackage{float}
\usepackage{color}
\usepackage{pstricks}
\usepackage{pst-node}
\usepackage{rotating}
\usepackage{overpic}
\usepackage{multirow}

\include{def-com}

\begin{document}
\normalsize
\parskip=5pt plus 1pt minus 1pt

\title{\boldmath {Evidence for $e^+e^-\to\gamma\chi_{c1, 2}$ at center-of-mass energies from 4.009 to 4.360 GeV }}

\input{authors_nov2014}

\begin{abstract}
Using data samples collected at center-of-mass energies of $\sqrt{s}$ = 4.009,
4.230, 4.260, and 4.360 GeV with the BESIII detector operating at the
BEPCII collider, we perform a search for the process
$e^+e^-\to\gamma\chi_{cJ}$ $(J = 0, 1, 2)$ and find evidence for
$e^+e^-\to\gamma\chi_{c1}$ and $e^+e^-\to\gamma\chi_{c2}$ with
statistical significances of 3.0$\sigma$ and 3.4$\sigma$,
respectively.   The Born cross sections
$\sigma^{B}(e^+e^-\to\gamma\chi_{cJ})$, as well as their upper limits
at the 90\% confidence level are determined at each center-of-mass
energy.
\end{abstract}

\pacs{14.40.Pq, 13.25.Gv, 13.66.Bc}
\maketitle
\section{Introduction} \label{sec::introduction}

The charmonium-like state $Y(4260)$ was first observed in the initial
state radiation (ISR) process $\EE\ar\gamma_{ISR}\ppjpsi$ by
BaBar~\cite{babar}, and later confirmed by the CLEO~\cite{cleo} and
Belle~\cite{belle} experiments. Recently, both BaBar and Belle
updated their results with full data sets, respectively,
and further confirmed the existence of the $Y(4260)$~\cite{babar2,
  belle2}.  Since it is produced through ISR in $\EE$ annihilation,
the $Y(4260)$ has the quantum numbers
$J^{PC} = 1^{--}$.  However, there seems to be no
$c\bar{c}$ slot available for the $Y(4260)$ in the conventional
charmonium
family~\cite{ref::slot}.  In addition, a number of unusual features,
such as a strong coupling to hidden-charm final states, suggest that
the $Y(4260)$ is a non-conventional $c\bar{c}$ meson. Possible
interpretations of this state can be found in Refs.~\cite{LQCD,
  tetra1, tetra2, mole1, mole2}, but all need further experimental
input.

Most previous studies of the $Y(4260)$ have utilized hadronic
transitions.  Besides the clear signal observed in the $\ppjpsi$ decay
mode, the Belle experiment failed to find evidence of the $Y(4260)$
via the $\EE\ar\gamma_{ISR}\eta\jpsi$ process~\cite{wangxl}.  Based on
13.2 $\rm pb^{-1}$ of $\EE$ data collected at $\sqrt{s}=4.260$~GeV, the CLEO
experiment investigated fourteen hadronic decay channels, but only few
decay modes had a significance more than 3$\sigma$~\cite{cleo2}. The
BESIII Collaboration first observed the process $e^+e^-\to\gamma X(3872)$
using data samples taken between $\sqrt{s} = 4.009$ and 4.420
GeV~\cite{ref::gammarec}, which strongly supports the existence of the
radiative transition decays of the $Y(4260)$.  To further understand
the nature of the $Y(4260)$ state, an investigation into the radiative
transitions between the $Y(4260)$ and other lower mass charmonium
states, like the $\chi_{cJ}$ $(J = 0, 1, 2)$, is important
~\cite{2013nna, 2014nna}.  The cross sections of
$e^+e^-\to\gamma\chi_{cJ}$ have been evaluated theoretically within
the framework of NRQCD~\cite{2014nna}.  Experimentally, the only
existing investigation comes from the CLEO experiment~\cite{cleo2},
which did not observe a signal.  The large data sample collected with the
BESIII detector provides a good opportunity to deeply investigate
these decay modes, which may shed more light on the properties of the
$Y(4260)$.

In this paper, we report on a search for $\EE\to\gamma\chi_{cJ}$
$(J=0, 1, 2)$ based on the large $\EE$ annihilation data samples
collected with the BESIII detector at center-of-mass energies (CME)
$\sqrt{s}=$ 4.009, 4.230, 4.260, and 4.360~GeV, where the $\chi_{cJ}$
is reconstructed by its $\gamma J/\psi$ decay mode, and the $J/\psi$
is by its decay to $\mu^+\mu^-$. The decay $J/\psi \to e^{+}e^{-}$ is not
considered in this analysis due to the high background of Bhabha
events.  The corresponding luminosities of the data samples at
different CME used in this analysis are listed in Table~\ref{tab::lum
  vs energy}.

\begin{table}[!htbp]
\begin{center}
\caption{The center-of-mass energy and Luminosity of each data sample.}
\label{tab::lum vs energy}
\begin{tabular}{cc}
\hline\hline ~~~~$\sqrt{s}$ (GeV)~~~~& ~~~~luminosity (pb$^{-1})$~~~~ \\ \hline
4.009  &   482  \\
4.230  &  1047  \\
4.260  &   826  \\
4.360  &   540  \\
\hline\hline
\end{tabular}
\end{center}
\end{table}

\section{BESIII detector and Monte Carlo} \label{sec::detector}
The BESIII detector at the BEPCII collider \cite{ref::bes3_detector}
is a large solid-angle magnetic spectrometer with a geometrical
acceptance of 93\% of $4\pi$ solid angle consisting of four main
components.  The innermost is a small-cell, helium-based (40\% He,
60\% C$_3$H$_8$) main drift chamber (MDC) with 43 layers providing an
average single-hit resolution of 135 $\mu$m.  The resulting
charged-particle momentum resolution for a 1~T magnetic field setting
is 0.5\% at 1.0 GeV$/c$, and the resolution on the ionization energy loss
information ($dE/dx$) is better than 6\%.  The next detector, moving
radially outwards, is a time-of-flight (TOF) system constructed of 5
cm thick plastic scintillators, with 176 detectors of 2.4 m length in
two layers in the barrel and 96 fan-shaped detectors in the end-caps.
The barrel (end-cap) time resolution of 80 ps (110 ps) provides a
$2\sigma$ $K/\pi$ separation for momenta up to 1.0 GeV.  Continuing
outward, we have an electromagnetic calorimeter (EMC) consisting of
6240 CsI(Tl) crystals in a cylindrical barrel structure and two
end-caps.  The energy resolution at 1.0 GeV is 2.5\% (5\%) and the
position resolution is 6 mm (9 mm) in the barrel (end-caps).  Finally,
the muon counter (MUC) consists of 1000 m$^2$ of Resistive Plate
Chambers (RPCs) in nine barrel and eight end-cap layers, which
provides a 2~cm position resolution.

A GEANT4~\cite{ref::geant4} based Monte Carlo (MC) simulation
software, which includes the geometric description
of the detector and the detector response, is used to optimize the
event selection criteria, determine the detection efficiency, and
estimate the potential backgrounds.  Signal MC samples of
$e^+e^-\to\gamma\chi_{cJ}$ are generated for each CME according to
the electric-diplole (E1) transition assumption~\cite{ref::evtgen}.
Effects of ISR are
simulated with KKMC~\cite{ref::kkmc} by assuming that $\gamma\chi_{cJ}$
is produced via $Y(4260)$ decays, where the $Y(4260)$ is described by a
Breit-Wigner function with resonance parameters from the Particle Data
Group (PDG)~\cite{ref::pdg2010}.
For the background studies, an `inclusive' $Y(4260)$ MC sample
equivalent to 500~$\rm pb^{-1}$ integrated luminosity is generated
which includes the $Y(4260)$ resonance, ISR production of the known
vector charmonium states, and events driven by QED processes. The
known decay modes are generated with EvtGen~\cite{ref::evtgen} with
branching fractions set to their world average values in the
PDG~\cite{ref::pdg2010}, and the remaining events are generated with
Lundcharm~\cite{ref::lundcharm} or PYTHIA~\cite{ref::pythia}.
%
\section{Event selection} \label{sec::selection}
%
%
Charged tracks are reconstructed in the MDC.   For each
good charged track, the polar angle must satisfy $|\cos\theta|<0.93$,
and the point of closest approach to the interaction point must be
within $\pm$10~cm in the beam direction and within $\pm$1 cm in the
plane perpendicular to the beam direction.
The number of good charged tracks is required to be
  two with a zero net charge.
Charged tracks are identified as muons if they have $E/p < 0.35$ and
$p>1.0$ GeV/c, where $E$ is the energy deposited in the EMC and $p$ is
the momentum measured by the MDC.

Photons are reconstructed from isolated showers in the EMC that are at
least 20 degrees away from any of the charged tracks. To improve the
reconstruction efficiency and the energy resolution, the energy
deposited in the nearby TOF counters is included. Photon candidates
are required to have energy greater than 25 MeV in the EMC barrel
region ($|\cos\theta|<0.8$), and 50 MeV in the end-cap region
($0.86<|\cos\theta|<0.92$).
In order to suppress electronic noise and energy deposits that are
unrelated to the event, the EMC time $t$ of the photon candidates must
be in coincidence with collision events within the range $0 \le t \le
700$ ns. At least two photon candidates in the final state are required.

\begin{figure*}[!htbp]
\centering
\subfigure[]{ \label{fig hl chic:subfig:a}

\includegraphics[width=0.43\textwidth]{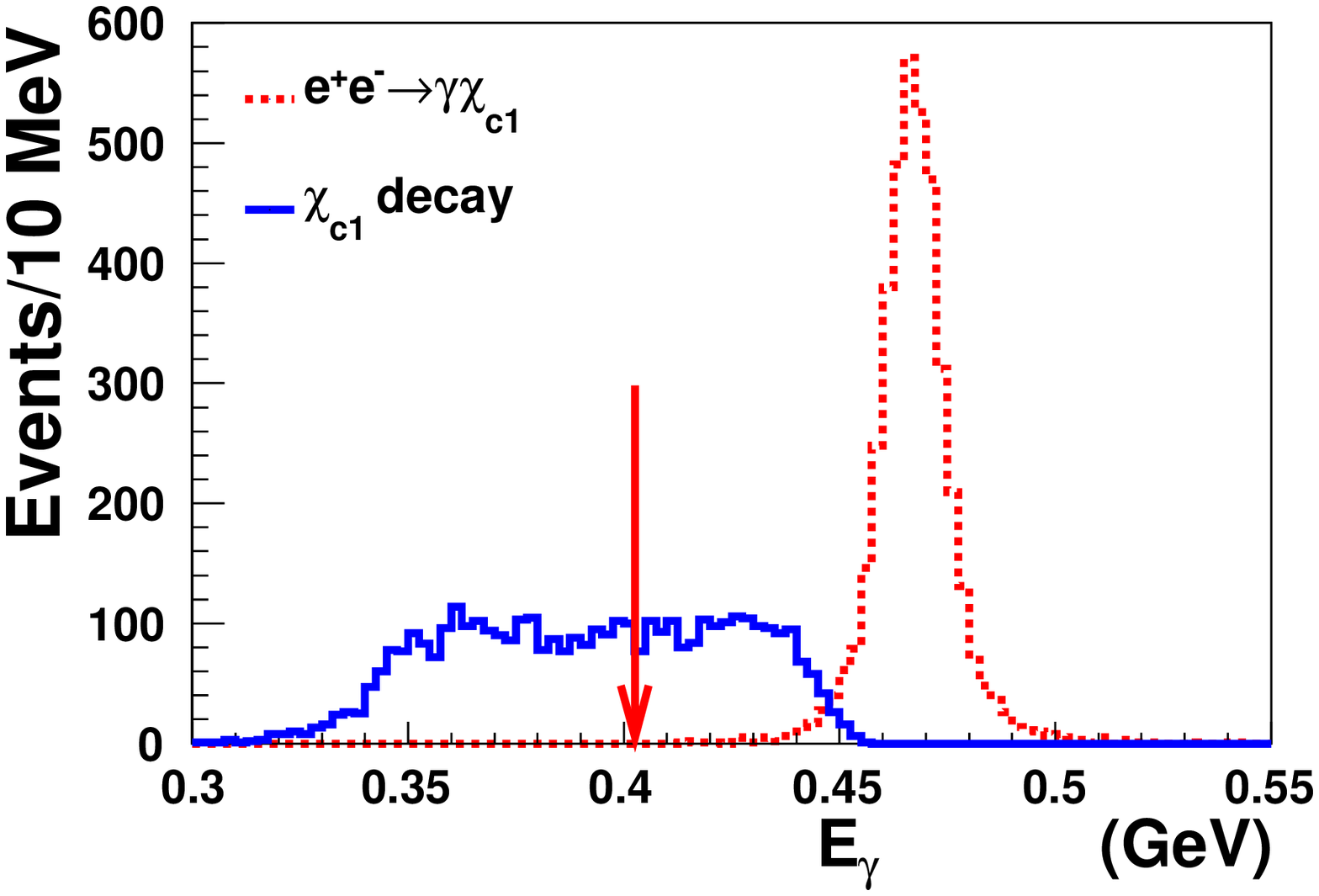}
}
\subfigure[]{ \label{fig hl chic:mini:subfig:b}
\includegraphics[width=0.43\textwidth]{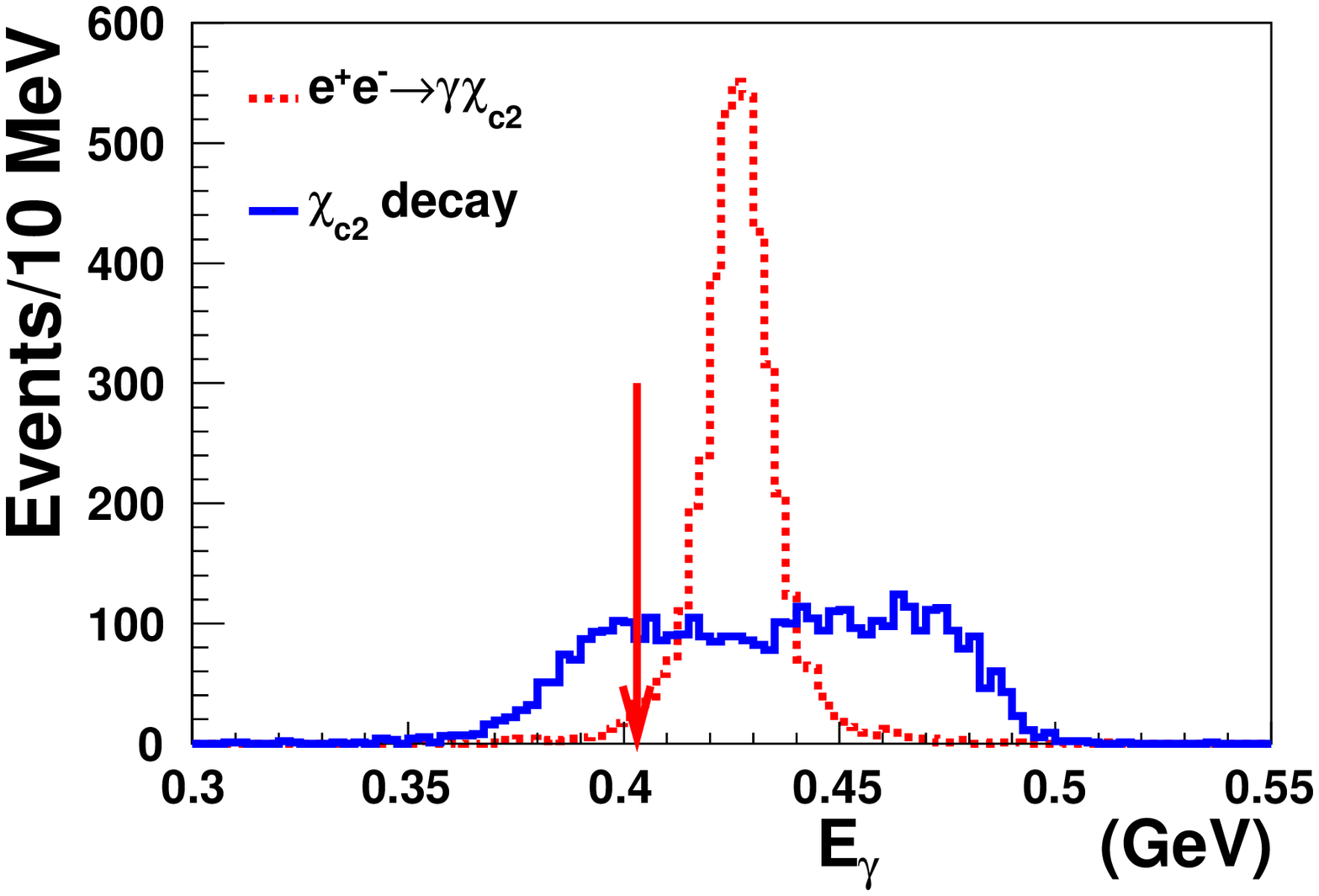}
}
\caption{ The distributions of photon energies in the laboratory frame
  from $e^+e^-\to\gamma\chi_{c1, 2}$ and from $\chi_{c1, 2}$ decays in
  the exclusive MC samples of $\EE\to\gamma\chi_{c1},
  \chi_{c1}\to\gamma J/\psi$ (a) and $\chi_{c2}$ (b) at $\sqrt{s}$ =
  4.009 GeV. Dashed lines stand for the first radiative photons from
  $e^+e^-\to\gamma\chi_{c1, 2}$ and solid lines for the second
  radiative photons from $\chi_{c1,2}$ decays.} \label{fig hl chic}
\end{figure*}

To improve the momentum resolution and to reduce backgrounds,
a kinematic fit with five constraints (5C-fit) is performed under the
$e^+e^-\ar\gamma\gamma\mu^+\mu^-$ hypothesis,
imposing overall energy and momentum conservation and constraining the
invariant mass of $\MM$ to the nominal $J/\psi$ mass.  Candidates with
a $\chi_{5C}^2 < 40$ are selected for further analysis.
%
If more than one candidate occurs in an event, the one with the
smallest $\chi_{5C}^2$ is selected.
Due to the kinematics of the reaction,
the first radiative photon from
$e^+e^-\to\gamma\chi_{cJ}$ has a high energy while the second
radiative photon from $\chi_{cJ}\to\gamma J/\psi$ has a lower energy
at $\sqrt{s}$ = 4.230, 4.260, and 4.360 GeV. The invariant mass of the low
energy photon and $\jpsi$, $M_{\gamma\jpsi}$, is used to search for
$\chi_{cJ}$ signals.
However, for the data sample taken at $\sqrt{s}=$4.009 GeV, there is an
overlap of the energy distributions of the photons from
$e^+e^-\to\gamma\chi_{c1, 2}$ and from $\chi_{c1, 2}$ decays, as
shown in Fig.~\ref{fig hl chic}. To separate the overlapping
photon spectra,
the energy of photons from $\chi_{c1, 2}$ decays is further required
to be less than 0.403~GeV at $\sqrt{s}=$4.009 GeV.

\section{Background study}
The potential backgrounds from $\EE\to P + J/\psi$, $P\ar\GG$
($P=\pi^0,\eta$, or $\etap$) can be rejected by requiring
$|M_{\gamma\gamma} - M_{\pi^0}| > 0.025$ GeV/$c^2$, $|M_{\gamma\gamma}
-M_{\eta}| > 0.03$ GeV/$c^2$ and $|M_{\gamma\gamma} - M_{\etap}| >
0.02$ GeV/$c^2$, where $M_{\gamma\gamma}$ is the invariant mass of two
selected photons.  The background from
$e^+e^-\to\gamma_{ISR}\psi(3686), \psi(3686)\to\gamma\chi_{cJ}$ is
rejected by applying the 5C kinematic fit.  After imposing all the
selection criteria above, the remaining dominant background is from
radiative dimuon events, which is not expected to peak in the
$M_{\gamma\jpsi}$ distribution.  This has been validated by
a dedicated simulation study.
For other remaining backgrounds, such as $\EE\to\pi^0\pi^0J/\psi$,
only few events (normalized to data luminosity) survive and can be
neglected.

\section{Fit to the mass spectrum}
\begin{figure*}[!htbp]
  \centering
  \subfigure[]{
  \label{fig gjpsi fit:mini:subfig:a}
  \includegraphics[width=0.43\textwidth]{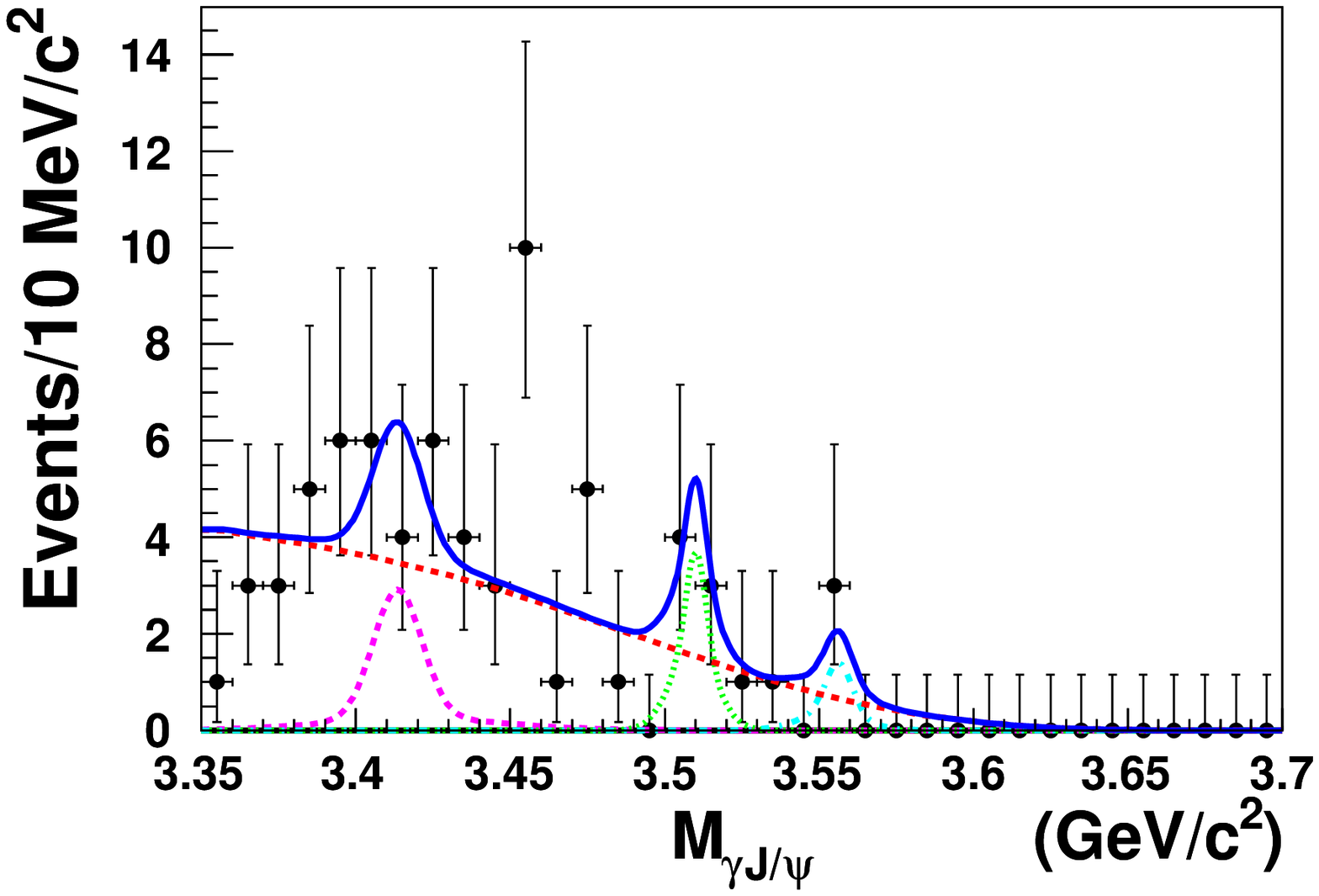}
  }
  \subfigure[]{
  \label{fig gjpsi fit:mini:subfig:b}
  \includegraphics[width=0.43\textwidth]{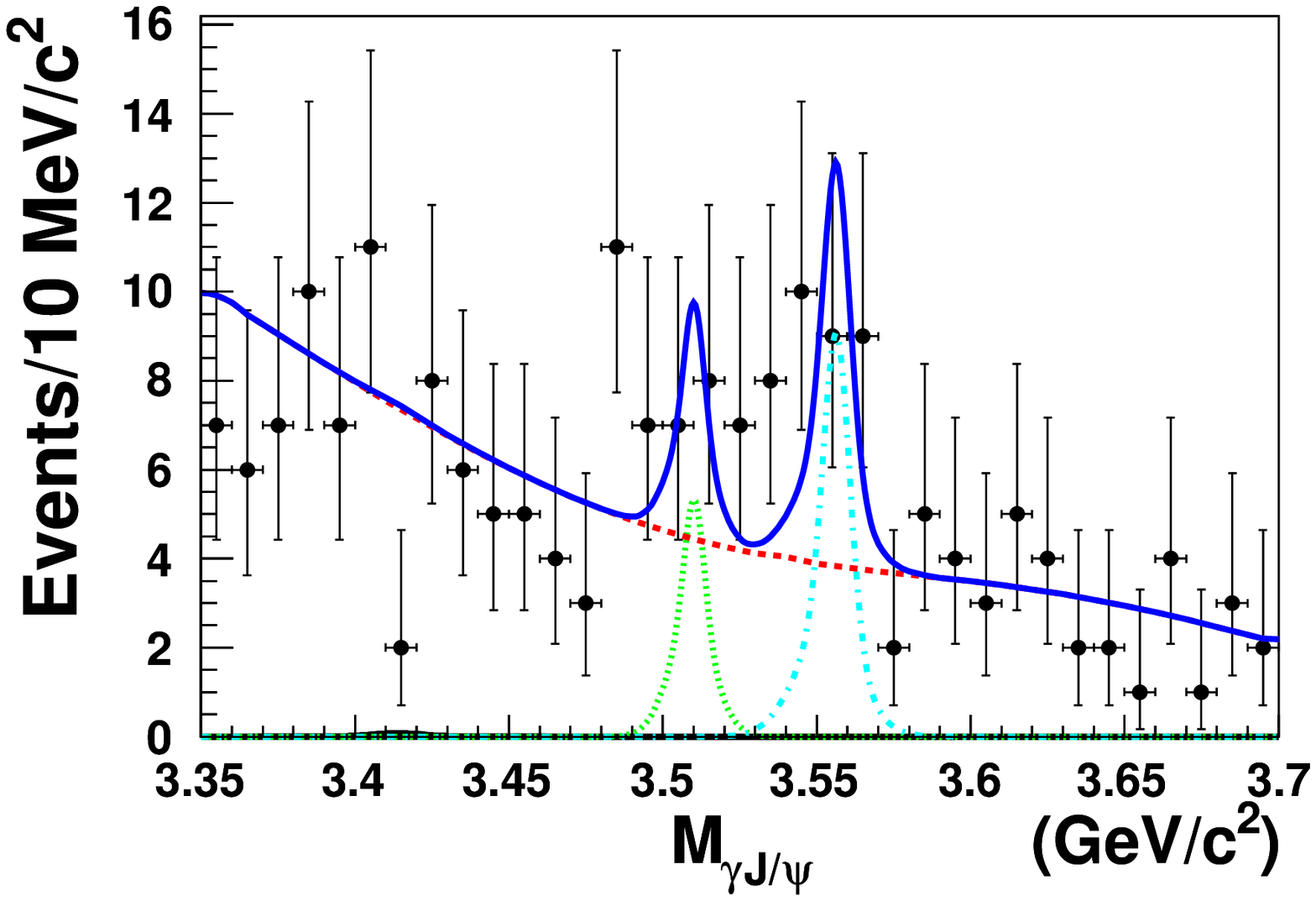}
  }
  \subfigure[]{
  \label{fig gjpsi fit:mini:subfig:c}
  \includegraphics[width=0.43\textwidth]{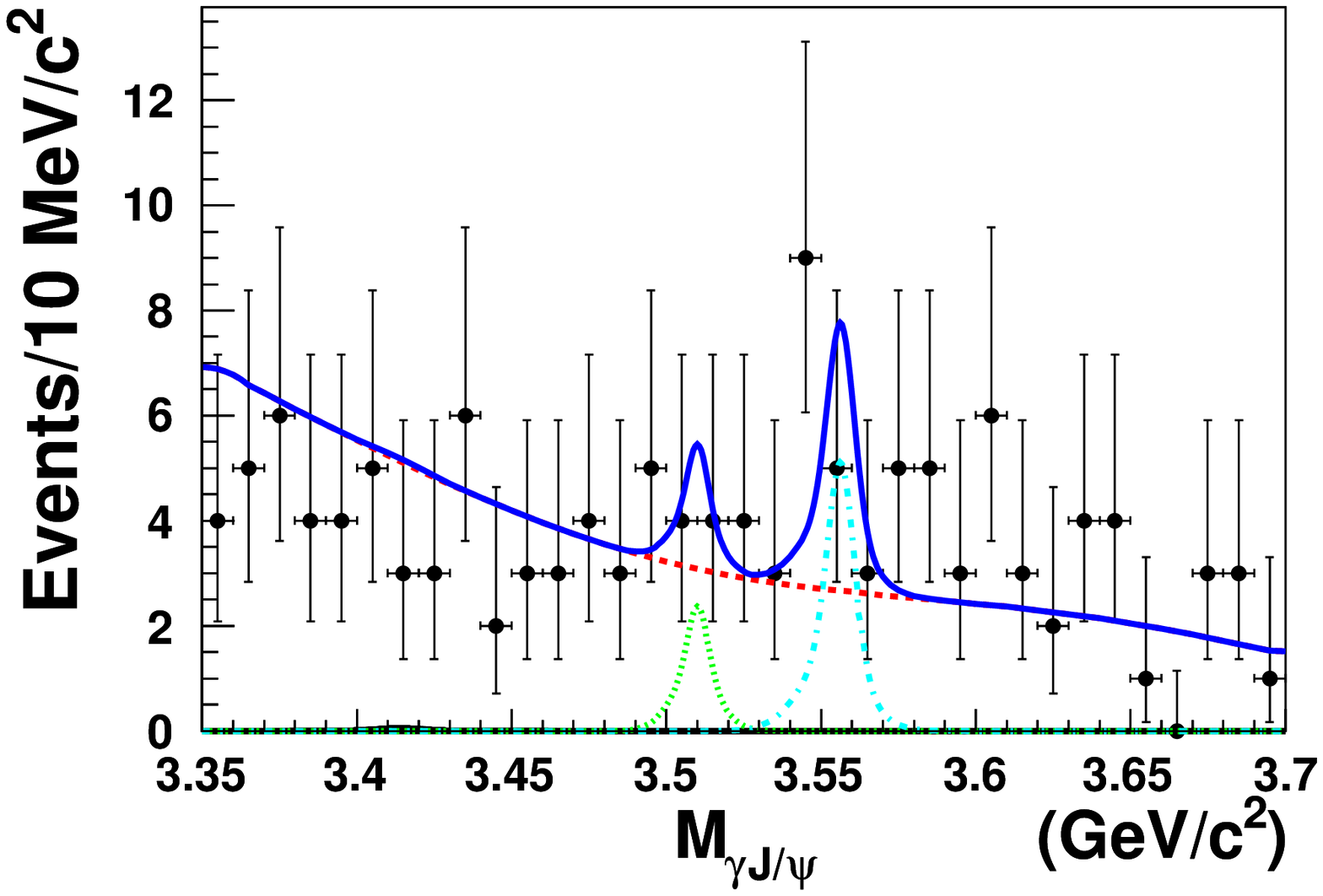}
  }
  \subfigure[]{
  \label{fig gjpsi fit:mini:subfig:d}
  \includegraphics[width=0.43\textwidth]{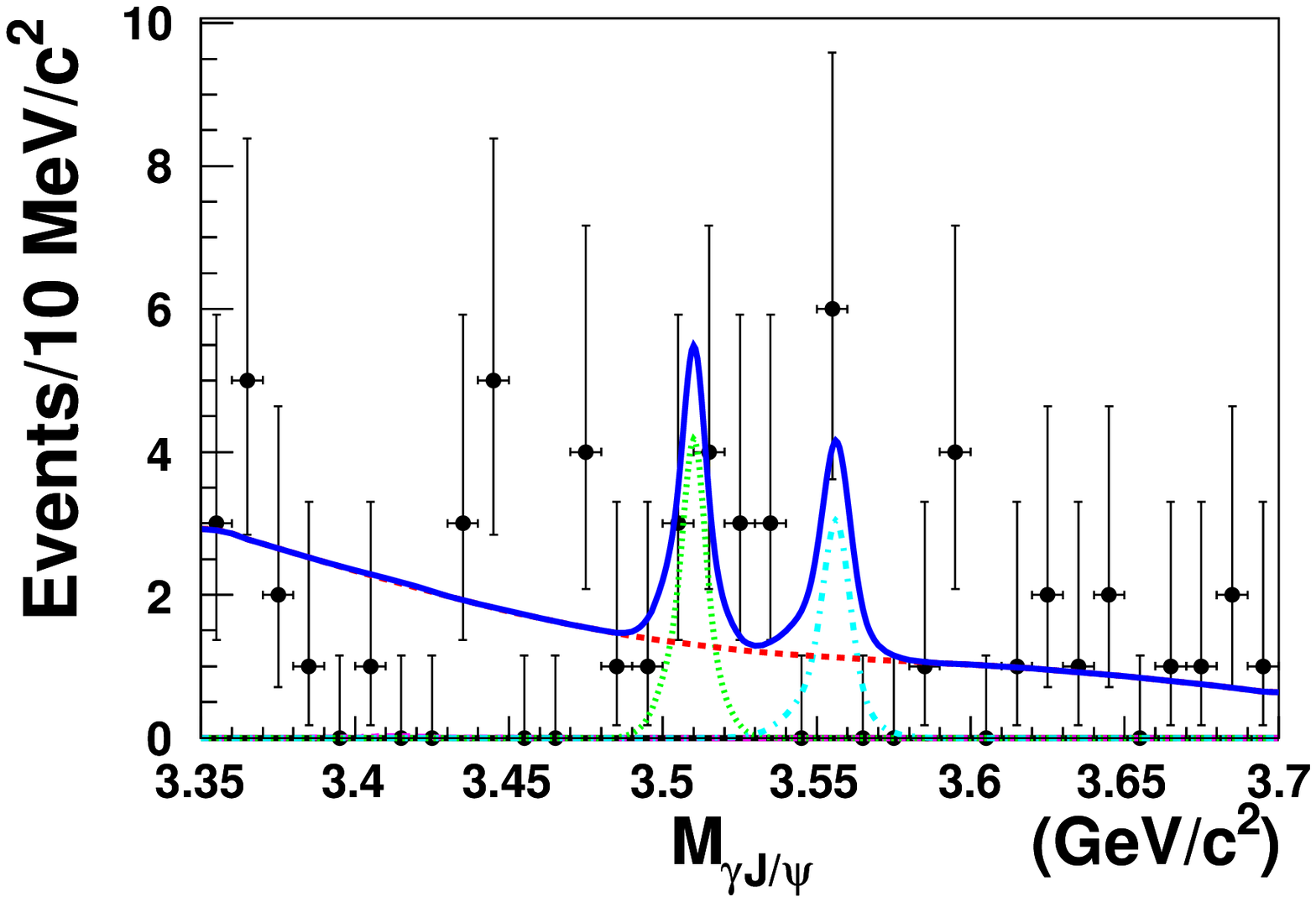}
  }
  \caption{The distribution of $\gamma J/\psi$ invariant mass,
    $M_{\gamma J/\psi}$, and fit results for data at
    $\sqrt{s}$ = 4.009 (a), 4.230 (b), 4.260 (c) and 4.360
    GeV  (d). The solid lines show the total fit results. The $\chicJ$
    signals are shown as dashed lines, dotted lines, and dash-dotted
    lines, respectively, and the backgrounds are indicated by red
    dashed lines.}
  \label{fig gjpsi fit}
  \end{figure*}

The resulting $M_{\gamma J/\psi}$ distributions, after applying the above selection criteria,
at $\sqrt{s}$ = 4.009, 4.230, 4.260 and 4.360 GeV are shown in
Figure~\ref{fig gjpsi fit}.  An unbinned maximum likelihood fit of the
$M_{\gamma J/\psi}$ distribution is performed to extract the numbers
of $\chi_{cJ}$ signal events. In the fit, the shapes of the $\chi_{cJ}$
signals are described by double Gaussian functions, where the means
and the standard deviations of the double Gaussian functions are
determined from a fit to the corresponding signal MC
  samples
at $\sqrt{s}$ = 4.260
GeV. These shapes are also used for the other three CME points, as the
resolution varies only mildly between $\sqrt{s}=4.009 - 4.360$ GeV.  This
has been validated by MC simulation. Since the dominant background
comes from radiative dimuon events, the corresponding MC simulation
is used to represent the background shape. To reduce the effect of statistical
fluctuations, the dimuon MC shape is smoothed before it is taken as the
background function. Figure~\ref{fig gjpsi fit} also shows the fitted
results for the $M_{\gamma\jpsi}$ distribution at different CME.  The
number of fitted $\chicJ$ signal events, as well as the corresponding
statistical significances (calculated by comparing the fit log
likelihood values with and without the $\chi_{cJ}$ signal) at the four
CME points are listed in Table~\ref{tab::ul}.
The same fit is applied to the
sum of $M_{\gamma J/\psi}$ distributions of the four
CME points.
The statistical significances for
$\chi_{c0}$, $\chi_{c1}$ and $\chi_{c2}$ are found to be 1.2$\sigma$, 3.0$\sigma$
and 3.4$\sigma$, respectively.  As a test, we perform similar
analyses to control samples from the $J/\psi$ sideband regions,
2.917 $< M_{\mu^+\mu^-} <$ 3.057 GeV/$c^2$ and 3.137 $< M_{\mu^+\mu^-}
<$ 3.277 GeV/$c^2$, by constraining the invariant mass of $\mu^+\mu^-$
to 3.047 or 3.147 GeV/$c^2$ in 5C-fit, and find no obvious $\chi_{cJ}$
signals.

\begin{table*}[!htbp]
\begin{center}
\renewcommand{\arraystretch}{1.2}
\setlength{\tabcolsep}{5pt}
\caption{The results on $e^+e^-\to\gamma\chi_{cJ}$ Born cross section
  measurement. Shown in the table are the significance $\sigma$,
  detection efficiency $\epsilon$, number of signal events from the
  fits N$^{\rm obs}$, radiative correction factor ($1+\delta^{r}$),
  vacuum polarization factor ($1+\delta^{v}$), upper limit (at the
  90\% C.L.) on the number of signal events N$^{\rm UP}$, Born cross
  section $\sigma^{B}$ and upper limit (at the 90\% C.L.) on the Born
  cross section $\sigma^{\rm UP}$ at different CME point. The first
  uncertainty of the Born cross section is statistical, and the second
  systematic.  }
\label{tab::ul}
\begin{tabular}{cccccccccc}
        \hline\hline
  $\sqrt{s}$ (GeV) & & N$^{\rm obs}$ &significance ($\sigma$)& N$^{\rm UP}$ & $\epsilon$ (\%) & $1+\delta^{r}$ & $1+\delta^{v}$

  & $\sigma^{\rm UP}$ (pb)&$\sigma^{B}$ (pb)\\ \hline

                \multirow{4}{*}{4.009} & $\chi_{c0}$ & 7.0$\pm$6.6 &1.6& 18 & 36.4$\pm$0.2&\multirow{4}{*}{0.738}&

                \multirow{4}{*}{1.04}&182&65.0$\pm$61.3$\pm$5.3\\

                & $\chi_{c1}$ & 4.4$\pm$2.6 &2.2& 9 & 23.4$\pm$0.1&&&5.3&2.4$\pm$1.4$\pm$0.2\\

                & $\chi_{c2}$ & 1.8$\pm$1.7 &1.5&  6 & 8.7$\pm$0.1&&&18&4.7$\pm$4.4$\pm$0.6\\ \hline

                \multirow{4}{*}{4.230} & $\chi_{c0}$ & 0.2$\pm$2.3 &0.0& 7 &

                37.2$\pm$0.2&\multirow{4}{*}{0.840}&\multirow{4}{*}{1.06}&26&0.7$\pm$8.0$\pm$0.1\\

                & $\chi_{c1}$ & 6.7$\pm$4.3 &1.9& 14 & 44.4$\pm$0.2&&&1.7&0.7$\pm$0.5$\pm$0.1\\

                & $\chi_{c2}$ & 13.3$\pm$5.2 &2.9& 22 & 42.0$\pm$0.2&&&5.0&2.7$\pm$1.1$\pm$0.3\\ \hline

                \multirow{4}{*}{4.260} & $\chi_{c0}$ & 0.1$\pm$1.9 &0.0& 5 &

                36.7$\pm$0.2&\multirow{4}{*}{0.842}&\multirow{4}{*}{1.06}&26&0.5$\pm$8.8$\pm$0.1\\

                & $\chi_{c1}$ & 3.0$\pm$3.0 &1.1& 7 & 42.7$\pm$0.2&&&1.1&0.4$\pm$0.4$\pm$0.1\\

                & $\chi_{c2}$ & 7.5$\pm$3.9 &2.3& 14 & 41.7$\pm$0.2&&&4.2&2.0$\pm$1.1$\pm$0.2\\ \hline

                \multirow{4}{*}{4.360} & $\chi_{c0}$ & 0.1$\pm$0.7 &0.0& 3 &

                32.4$\pm$0.2&\multirow{4}{*}{0.943}&\multirow{4}{*}{1.05}&23&0.7$\pm$5.0$\pm$0.1\\

                & $\chi_{c1}$ & 5.2$\pm$4.9 &2.4& 10 & 31.7$\pm$0.2&&&2.9&1.4$\pm$1.3$\pm$0.1\\

                & $\chi_{c2}$ & 4.4$\pm$4.5 &2.0& 9 & 30.3$\pm$0.2&&&5.0&2.3$\pm$2.3$\pm$0.2\\
        \hline \hline
\end{tabular}
\end{center}
\end{table*}

\begin{table*}[!htbp]
\begin{center}
\renewcommand{\arraystretch}{1.2}
\caption{Summary of systematic uncertainties at $\sqrt{s}$ = 4.009, 4.230, 4.260, and 4.360 GeV(\%).} \label{tab::systematic}
\begin{tabular}{c c c c c c c c c c c c c c c c c c}
\hline \hline
$\sqrt{s}$ (GeV)& & &4.009 & & & &4.230 & & & &4.260 & & & &4.360\\
Sources & &$\chi_{c0}$&$\chi_{c1}$&$\chi_{c2}$ & & $\chi_{c0}$&$\chi_{c1}$&$\chi_{c2}$& & $\chi_{c0}$&$\chi_{c1}$&$\chi_{c2}$& & $\chi_{c0}$&$\chi_{c1}$&$\chi_{c2}$\\ \hline
Luminosity        &  & 1.0  & 1.0 & 1.0 &  & 1.0  & 1.0 & 1.0  &  & 1.0  & 1.0 & 1.0 &  & 1.0  & 1.0 & 1.0\\
Tracking efficiency & & 2.0  & 2.0 & 2.0 &  & 2.0  & 2.0 & 2.0  &  & 2.0  & 2.0 & 2.0 &  & 2.0  & 2.0 & 2.0\\
Photon detection    & & 2.0  & 2.0 & 2.0 &  & 2.0  & 2.0 & 2.0  &  & 2.0  & 2.0 & 2.0 &  & 2.0  & 2.0 & 2.0\\
Kinematic fit   & & 0.6  & 0.6 & 0.6 &  & 0.6  & 0.6 & 0.6  &  & 0.6  & 0.6 & 0.6 &  & 0.6  & 0.6 & 0.6\\
Branching ratio   & & 4.8  & 3.6 & 3.7 &  & 4.8  & 3.6 & 3.7  &  & 4.8  & 3.6 & 3.7 &  & 4.8  & 3.6 & 3.7\\
Vacuum polarization factor   & & 0.5  & 0.5 & 0.5 &  & 0.5  & 0.5 & 0.5  &  & 0.5  & 0.5 & 0.5 &  & 0.5  & 0.5 & 0.5\\
$\chi_{cJ}$ mass resolution     &  &0.3   & 2.0 & 7.4 &  & 0.0  & 7.7 & 7.8  &  & 0.0  & 4.3 & 6.5 &  & 0.0  & 1.1 & 2.0\\
$\chi_{cJ}$ mass     & &0.0   & 0.9 & 1.4 &  & 0.0  & 0.3 & 0.2  &  & 0.0  & 0.1 & 0.1 &  & 0.0  & 0.3 & 0.4\\
Decay mode        & & 4.9  & 2.2 & 3.9 &  & 5.5  & 1.2 & 3.3  &  & 5.9  & 1.9 & 2.9 &  & 5.1  & 1.5 & 2.1\\
Fit range       & & 0.1  & 2.2 & 2.6 &  & 0.0  & 1.5 & 2.3  &  & 0.0  & 3.1 & 2.5 &  & 0.0  & 3.0 & 3.7\\
Background shape      &  & 0.0  & 3.1 & 5.6 &  & 0.0  & 0.7 & 0.3  &  & 0.0  & 1.1 & 0.4 &  & 0.0  & 0.9 & 0.1\\
Radiative correction factor   & & 3.0  & 2.7 & 3.6 &  & 2.6  & 3.1 & 2.1  &  & 3.5  & 2.1 & 2.5 &  & 1.8  & 3.4 & 3.8\\
\hline
Total            & & 8.1 &7.3  & 12.1 & & 8.3 &9.8  & 10.2 & & 8.9 &7.7  & 9.3 & & 7.9 &6.9  & 7.7\\
\hline\hline
\end{tabular}
\end{center}
\end{table*}
\section{Results}
The Born cross section at different CME is calculated with
\begin{equation}
\sigma^{B}(e^+e^-\to\gamma\chi_{cJ}) = \frac{N^{\rm obs}}{\mathcal{L}\cdot(1+\delta^{r})\cdot(1+\delta^{v})\cdot\mathcal{B}\cdot\epsilon}
\end{equation}
where $N^{\rm obs}$ is the number of observed events obtained from the
fit, $\mathcal{L}$ is the integrated luminosity, $1+\delta^{r}$ is the
radiative correction factor for $\chi_{cJ}$ with the assumption that
the $e^+e^-\to\gamma\chi_{cJ}$ cross section follows the $Y(4260)$
Breit-Wigner line shape~\cite{ref::radiative}, $1+\delta^{v}$ is the vacuum polarization
factor~\cite{ref::vacuum}, $\mathcal{B}$ is the combined branching ratio of $\chi_{cJ}
\to \gamma J/\psi$ and $J/\psi\to\mu^+\mu^-$, and $\epsilon$ is the
detection efficiency.  The detection efficiencies, radiative
correction factors as well as the calculated Born cross sections at
different CME are shown in Table~\ref{tab::ul}. The much lower
efficiencies for $\chi_{c1,2}$ at $\sqrt{s}$ = 4.009 GeV are due to
the requirement on the photon energy used to separate the overlapping
photon spectra as described in Section~\ref{sec::selection}.

Since the $\chi_{cJ}$ signals are not statistically significant at
the individual CME points,
we also give in Table~\ref{tab::ul} the
 by assuming the non-existence of signals, the upper limits on the
 Born cross sections at the 90\% confidence level (C. L.) under the
 assumption that no signals are present.  The upper limits are derived
using a Bayesian method~\cite{ref::pdg2010}, where the efficiencies
are lowered by a factor of $(1 - \sigma_{\rm sys})$ to take
systematic uncertainties into account.

\begin{figure}[!htbp]
\begin{center}
\includegraphics[width=0.45\textwidth]{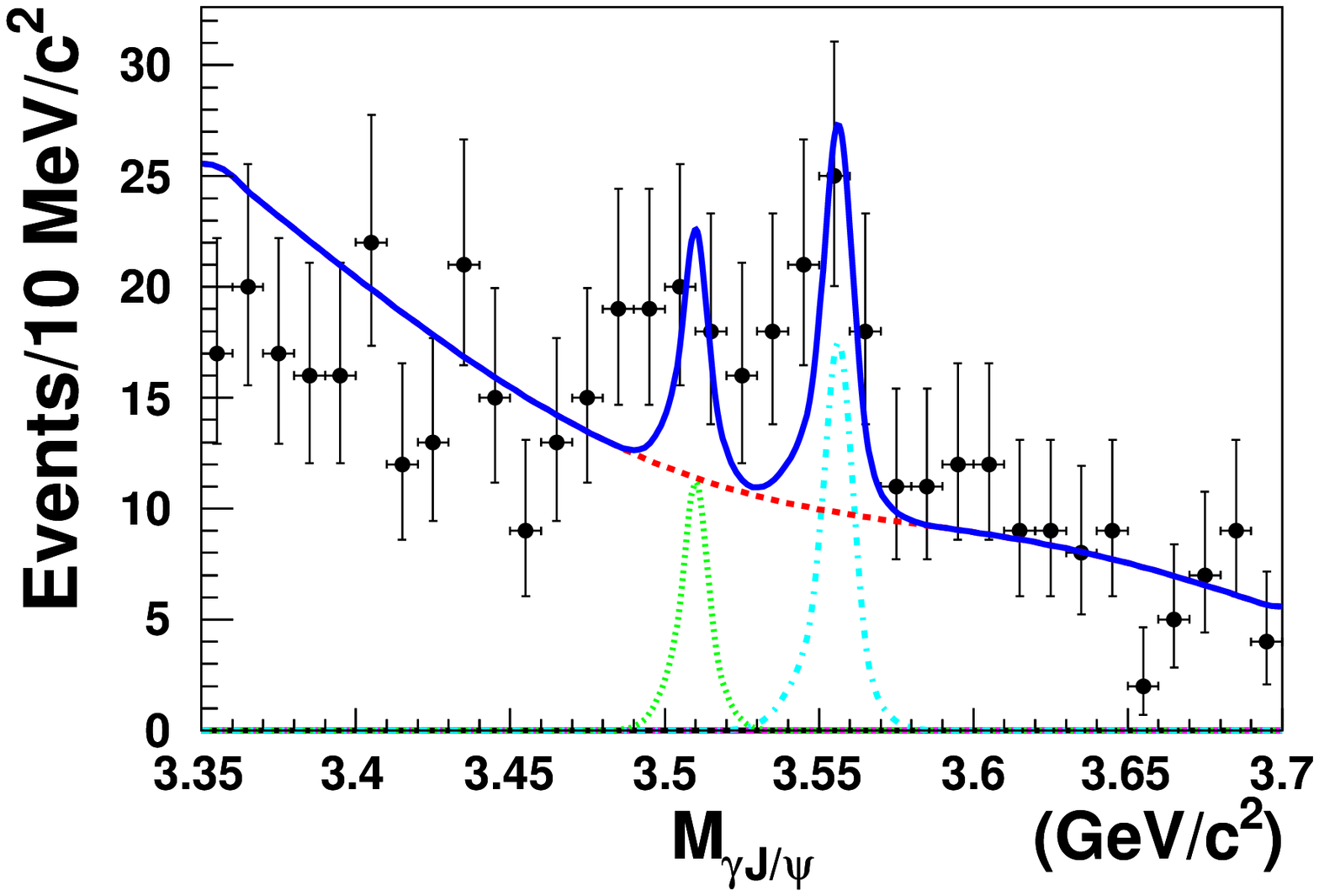}
\caption{Result of the simultaneous fit to
  $M_{\gamma\jpsi}$ distributions for all CME data
  sets assuming that the signals are from decays of the $Y(4260)$. The
  blue solid line is the total fit result. The $\chicJ$ signals are shown as dashed
  line, dotted line, and dash-dotted line, respectively, and the
  background is shown as the red dashed line.}
\label{fig fit y4260 shape}
\end{center}
\end{figure}

We also perform a simultaneous fit to the $M_{\gamma J/\psi}$ distribution
at $\sqrt{s}$ = 4.009, 4.230, 4.260, and 4.360 GeV, assuming the
production cross section of $\EE\to\gamma\chi_{cJ}$ at different CME
point follows the line shape of the Y(4260) state.
In the fit, the line shapes of the $\chi_{cJ}$ signals and the background
are as same as those in previous fits, and the number of $\chi_{cJ}$
events at each CME point is expressed as a function of $\epsilon_{\rm
  c.m.}\mathcal{L}_{\rm c.m.}\rm R_{\rm c.m.}(1+\delta^{r})$, where
$\epsilon_{\rm c.m.}$ and $\mathcal{L}_{\rm c.m.}$ are the detection
efficiency and luminosity, respectively, and $\rm R_{c.m.}$ is the
ratio of the cross section calculated with the $Y(4260)$ line shape (a
Breit-Wigner function with parameters fixed to the PDG values) at
different CME points to that at $\sqrt{s}$ = 4.260 GeV.  The
corresponding fit result is shown in Fig. ~\ref{fig fit y4260 shape},
and the statistical significances for $\chi_{c0}$, $\chi_{c1}$ and
$\chi_{c2}$ signals are 0.0$\sigma$, 2.4$\sigma$ and 4.0$\sigma$,
respectively.

\section{Systematic uncertainties} \label{subsec::systematic}
The systematic uncertainties in the cross section measurements of
$\EE\ar\gamma\chi_{cJ}$ caused by various sources are partially in
common for all channels. The common sources of systematics include the
luminosity measurement, reconstruction efficiencies
  for charged tracks and photons,
the vacuum polarization factor, kinematic fit and branching fractions of
the decay of the intermediate states.  The systematic uncertainty due
to the luminosity measurement is estimated to be 1.0\% using Bhabha
events.  The uncertainty related to the track reconstruction
efficiency of high-momentum muons is 1.0\% per track
\cite{ref::muontrack}.  The systematic uncertainty related to the
photon detection is estimated to be 1.0\% per photon
\cite{ref::gammarec}.  The systematic uncertainty due to 5C-fit is
0.6\%, obtained by studying a control sample of $\psp\ar\eta\jpsi$
decays.  The uncertainty related to the branching fractions of
$\chi_{cJ}$ and $\jpsi$ decays are taken from the PDG~\cite{ref::pdg2010}.
The uncertainty for the vacuum polarization factor is
0.5\%~\cite{ref::vacuum}.

The other systematic uncertainties arising from the $\chi_{cJ}$ mass
resolution, the shift of the $\chi_{cJ}$ reconstructed mass, the MC model,
the shape of background, the radiative correction factor and the fit
range at different CME points are discussed below.

The $\psp\to\gamma\chi_{cJ}$ channel is employed as a control sample
to extract the differences on the mass resolution of the $\chi_{cJ}$
signal by fitting the $M_{\gamma\jpsi}$ spectrum. The differences in
the mass resolutions between data and MC are found to be 0.02\%,
0.01\%, 0.2\% for $\chi_{cJ}$ $(J = 0, 1, 2)$. A similar fit
is performed, in which
the signal shapes are smeared  to compensate for the mass resolution
difference,
and the differences on the yields of $\chicJ$
signal are taken as the systematic uncertainties due to the mass
resolution.

An alternative fit is performed shifting the mean of $\chi_{cJ}$
signal shapes by one standard deviation of the PDG values, and the
deviations of the signal yields to the nominal values are taken as
the systematic uncertainties due to the uncertainties of the signal
line shapes.

The detection efficiency is evaluated using MC samples based on the E1
transition assumption~\cite{ref::evtgen} for
$Y(4260)\to\gamma \chi_{cJ}$.
Another set of MC samples
is generated where the $Y(4260)\to\gamma \chi_{cJ}$ decay
is modeled using a phase space distribution,
and the differences of the detector efficiencies
between the two sets of MC samples are treated as systematic
uncertainties from the MC model.

To estimate the systematic uncertainty related to the background
shape, a control sample is selected from the data by requiring a
$\mu^+\mu^-$ pair and at least one photon. An alternative background
shape is then extracted by re-weighting the $\gamma\mu^+\mu^-$
invariant mass spectrum of the control sample, where the weights are
the efficiency ratio of $\EE\to(n\gamma)\mu^+\mu^-$ MC simulated
events surviving the signal selection criteria to the same selection
criteria for the control sample. A fit with the alternative
background shape is performed, and the differences between the yields
of $\chi_{cJ}$ signal to the nominal ones are taken as the systematic
uncertainties due to the shape of background.

The possible distortions of the $Y(4260)$ line shape due to
interference effects with nearby resonances could introduce
uncertainties in the radiative correction factor
$\epsilon\times(1+\delta^{r})$. To estimate the related systematic
uncertainties, we instead assume that $e^+e^-\to\gamma\chi_{cJ}$ are
produced via $\psi(4040)$ decays at $\sqrt{s}$ = 4.009 GeV,
$\psi(4160)$ decays at $\sqrt{s}$ = 4.229 and 4.260 GeV, and
$\psi(4415)$ decays at $\sqrt{s}$ = 4.360 GeV. The variations in the
factor $\epsilon\times(1+\delta^{r})$ are taken as the systematic
uncertainties due to the radiative correction factor.

A series of similar fits are performed in different ranges of the
$M_{\gamma J/\psi}$ distribution, and the largest differences on the
signal yields to the nominal values are taken as systematic
uncertainties.

All the systematic uncertainties from the different sources are summarized
in Table~\ref{tab::systematic}. The total systematic uncertainties are
calculated as the quadratic sum of all individual terms.

\section{Summary} \label{sec::summary}
Using data samples collected at CME of $\sqrt{s}$ = 4.009, 4.230,
4.260, and 4.360 GeV with the BESIII detector, we perform a search
for $e^+e^-\to\gamma\chi_{cJ}$ $(J = 0, 1, 2)$ with the subsequent
decay $\chi_{cJ}\to\gamma J/\psi$ and $J/\psi \to \mu^+\mu^-$.  We
find evidence for the processes $e^+e^-\to\gamma\chi_{c1}$ and
$e^+e^-\to\gamma\chi_{c2}$ with statistical significances of
3.0$\sigma$ and 3.4$\sigma$, respectively. No evidence of
$\EE\to\gamma\chi_{c0}$ is observed.  The corresponding Born
cross sections of $e^+e^-\to\gamma\chi_{cJ}$ at different CME are
calculated and listed in Table~\ref{tab::ul}.
Under the assumption of the absence of $\chi_{cJ}$ signals,
the upper limits on the Born
cross sections at the 90\% C.L. are calculated and listed in Table
\ref{tab::ul}, too.  These upper limits on the Born cross section of
$\EE\to\gamma\chi_{cJ}$ are compatible with the theoretical prediction
from an NRQCD calculation~\cite{2014nna}.

\section{Acknowledgement}
The BESIII collaboration thanks the staff of BEPCII and the IHEP
computing center for their strong support. This work is supported in
part by National Key Basic Research Program of China under Contract
No.~2015CB856700; Joint Funds of the National Natural Science
Foundation of China under Contracts Nos.~11079008, 11179007, U1232201,
U1332201; National Natural Science Foundation of China (NSFC) under
Contracts Nos.~10935007, 11121092, 11125525, 11235011, 11322544,
11335008; the Chinese Academy of Sciences (CAS) Large-Scale Scientific
Facility Program; CAS under Contracts Nos.~KJCX2-YW-N29, KJCX2-YW-N45;
100 Talents Program of CAS; INPAC and Shanghai Key Laboratory for
Particle Physics and Cosmology; German Research Foundation DFG under
Contract No.\ Collaborative Research Center CRC-1044; Istituto
Nazionale di Fisica Nucleare, Italy; Ministry of Development of Turkey
under Contract No. DPT2006K-120470; Russian Foundation for Basic
Research under Contract No. 14-07-91152; U. S. Department of Energy
under Contracts Nos.\ DE-FG02-04ER41291, DE-FG02-05ER41374,
DE-FG02-94ER40823, DESC0010118; U.S. National Science Foundation;
University of Groningen (RuG) and the Helmholtzzentrum fuer
Schwerionenforschung GmbH (GSI), Darmstadt; WCU Program of National
Research Foundation of Korea under Contract No.\ R32-2008-000-10155-0.


\end{document}

%% file: def-com.tex
\newcommand{\chicJ}{\chi_{cJ}}

\newcommand{\EE}{e^+e^-}
\newcommand{\MM}{\mu^+\mu^-}
\newcommand{\GG}{\gamma\gamma}

\newcommand{\etap}{\eta^{\prime}}

\newcommand{\psp}{\psi^{\prime}}
\newcommand{\jpsi}{J/\psi}
\newcommand{\ar}{\rightarrow}

\newcommand{\ppjpsi}{\pi^+\pi^- J/\psi}

\newcommand{\bfg}{\begin{figure}}
\newcommand{\efg}{\end{figure}}
\newcommand{\bitm}{\begin{itemize}}
\newcommand{\eitm}{\end{itemize}}
\newcommand{\bnum}{\begin{enumerate}}
\newcommand{\enum}{\end{enumerate}}
\newcommand{\btbl}{\begin{table}}
\newcommand{\etbl}{\end{table}}
\newcommand{\btbu}{\begin{tabular}}
\newcommand{\etbu}{\end{tabular}}
\newcommand{\bcl}{\begin{center}}
\newcommand{\ecl}{\end{center}}
\newcommand{\bbt}{\bibitem}
\newcommand{\beq}{\begin{equation}}
\newcommand{\eeq}{\end{equation}}
\newcommand{\beqr}{\begin{eqnarray}}
\newcommand{\eeqr}{\end{eqnarray}}

%% file: authors_nov2014.tex
\author{
  \begin{small}
    \begin{center}
      M.~Ablikim$^{1}$, M.~N.~Achasov$^{8,a}$, X.~C.~Ai$^{1}$,
      O.~Albayrak$^{4}$, M.~Albrecht$^{3}$, D.~J.~Ambrose$^{43}$,
      A.~Amoroso$^{47A,47C}$, F.~F.~An$^{1}$, Q.~An$^{44}$,
      J.~Z.~Bai$^{1}$
% ...
      \\
      \vspace{0.2cm}
      (BESIII Collaboration)\\
      \vspace{0.2cm} {\it
        $^{1}$ Institute of High Energy Physics, Beijing 100049, People's Republic of China\\
% ...
$^{3}$ Bochum Ruhr-University, D-44780 Bochum, Germany\\
        $^{4}$ Carnegie Mellon University, Pittsburgh, Pennsylvania 15213, USA\\
% ...
$^{8}$ G.I. Budker Institute of Nuclear Physics SB RAS (BINP), Novosibirsk 630090, Russia\\
% ...
$^{43}$ University of Rochester, Rochester, New York 14627, USA\\
        $^{44}$ University of Science and Technology of China, Hefei 230026, People's Republic of China\\
% ...
$^{47}$ (A)University of Turin, I-10125, Turin, Italy; (B)University of Eastern Piedmont, I-15121, Alessandria, Italy; (C)INFN, I-10125, Turin, Italy\\
% ...
$^{a}$ Also at the Novosibirsk State University, Novosibirsk, 630090, Russia\\
% ...
      }\end{center}
    \vspace{0.4cm}
  \end{small}
}

\affiliation{}